\documentclass[reprint,aps,prl,twocolumn,superscriptaddress,floatfix,amsmath,amssymb,10pt]{revtex4-2}
\DeclareMathAlphabet{\mathpzc}{OT1}{pzc}{m}{it}
\usepackage{bm}
\usepackage{dcolumn}
\usepackage{amsthm}
\usepackage{amsmath}
\usepackage{amssymb}
\usepackage{graphicx}
\usepackage{xcolor}
\usepackage{fix-cm}
\usepackage{mathptmx} 
\usepackage[T1]{fontenc}
\usepackage[colorlinks,allcolors=blue]{hyperref}
\makeatletter
\def\NAT@def@citea{\def\@citea{\NAT@separator}}
\makeatother

\begin{document}

\title{Theoretical Uncertainty Quantification for Heavy-ion Fusion}

\author{K. Godbey}\email{science@kyle.ee}
\affiliation{Facility for Rare Isotope Beams, Michigan State University, East Lansing, Michigan 48824, USA }
\author{A.S. Umar}\email{umar@compsci.cas.vanderbilt.edu}
\affiliation{Department of Physics and Astronomy, Vanderbilt University, Nashville, Tennessee 37235, USA}
\author{C. Simenel}\email{cedric.simenel@anu.edu.au}
\affiliation{Department of Fundamental and Theoretical Physics, Research School of Physics,
The Australian National University, Canberra ACT 2601, Australia}

\date{\today}


\begin{abstract}
Despite recent advances and focus on rigorous uncertainty quantification for microscopic models of quantum many-body systems, the uncertainty on the dynamics of those systems has been under-explored.
To address this, we have used time-dependent Hartree-Fock to examine the model uncertainty for a collection of low-energy, heavy-ion fusion reactions.
Fusion reactions at near-barrier energies represent a rich test-bed for the dynamics of quantum many-body systems owing to the complex interplay of collective excitation, transfer, and static effects that determine the fusion probability of a given system.
While the model uncertainty is sizable for many of the systems studied, the primary contribution comes from ill-constrained static properties, such as the neutron radius of neutron-rich nuclei.
These large uncertainties motivate the use of information from reactions to better constrain existing models and to infer static properties from reaction data.
\end{abstract}
\maketitle

\paragraph{Introduction.} A robust description of the dynamics of atomic nuclei is at the heart of understanding many aspects of physics that span from the origin of the elements to the complex landscape on the surface of neutron stars.
Nucleosynthesis, for instance, involves fusion of light nuclei, transfer and formation of neutron-rich nuclei, and the fission and subsequent recycling of fission products of heavy nuclei.
Although reactions with stable nuclei have been extensively studied experimentally in the past, those with neutron rich nuclei that exhibit exotic structures such as neutron skins and haloes are less known. 
Predictive theoretical models of low-energy nuclear dynamics are then crucial for reliable descriptions of such reactions. 
In addition, the recent development of exotic beams has significantly increased the
range of available systems for reaction studies that could test these models, creating exciting opportunities
 at exotic beam facilities around the
world, including FRIB (US)~\cite{glasmacher2017}, RIKEN-RIBF
(Japan)~\cite{sakurai2010}, SPIRAL2 (France)~\cite{lewitowicz2011}, and
FAIR-NUSTAR (Germany)~\cite{kalantar2018}.

While there are many theoretical approaches to studying these disparate physical processes commonly encountered in the
 study of nuclei, it is advantageous to utilize a framework that can be more consistently employed to a wide swath of these problems.
Time-dependent Hartree-Fock (TDHF) theory is one such framework \cite{negele1982}. %
Although it has been applied to studies of various low energy heavy-ion reactions including multi-nucleon transfer, deep-inelastic collisions, quasi-fission and fission,  
one of the most important applications is the study of fusion (see ~\cite{simenel2012,simenel2018,stevenson2019,sekizawa2019} for recent reviews). 
The latter is particularly interesting as it occurs over a broad range of energies from well below to well above the Coulomb barrier between the reactants, resulting in fusion cross-sections spanning many orders of magnitude. 
As a result, fusion magnifies quantum effects such as tunneling and coherent couplings between relative motion and internal degrees of freedom of the collision partners \cite{back2014}. Low-energy fusion is therefore very sensitive to the structure of the reactants.
Naturally, microscopic approaches such as TDHF that describe nuclear structure and dynamics on the same footing are well suited to investigate such reactions. 

At the heart of TDHF is the energy density functional (EDF) which encodes the nucleon-nucleon interaction as a functional of various nuclear densities coupled by a set of parameters that solely defines the properties of the nuclear system and its dynamics.
Despite the development of many EDFs over the years, it is only recently that a rigorous Bayesian uncertainty quantification study has been performed for EDF parameter estimation~\cite{mcdonnell2015}.
Thus, while the extent to which nuclear dynamics are sensitive to varying terms in the EDF has been explored before~\cite{reinhard2016a}, a similar study within the uncertainties of a fully quantified model has remained elusive.

The purpose of this work is to investigate the uncertainties, originating from the EDF, in the prediction of fusion cross-sections from deep sub-barrier to above barrier energies. 
Despite well characterized methods and impressive results, the extent to which these time-dependent microscopic model predictions are uncertain as a result of the model parameter determination itself is unknown.
It is to this end that we have systematically studied model predictions from an ensemble of EDFs that have been sampled from the Bayesian posterior distributions.
In this letter, the density-constrained TDHF method is used to study the fusion of the $^{40,48}$Ca$+^{40,48}$Ca and $^{16}$O$+^{208}$Pb systems.
The results are compared to available experimental data and the potential impact of fusion reactions on EDF development are explored.

\paragraph{Theoretical methods.} Historically,
direct TDHF studies were used to study fusion above the barrier by determining the maximum
impact parameter value that results in fusion. The assumption of a sharp cutoff 
(fusion occurs below the sharp cutoff value with probability one) in impact
parameter was then used to calculate the fusion cross-sections. 
However, the lack of many-body
tunneling in real-time dynamical mean-field approaches prevented the direct calculation of sub-barrier fusion cross-sections with TDHF. 
Although extensions of TDHF to imaginary-time accounts for tunneling at the mean-field level \cite{levit1980c}, 
applications are limited to simple systems \cite{mcglynn2020}. 

An alternative approach to account for tunneling is provided by  the so called density constrained TDHF method (DC-TDHF)~\cite{umar2006b}. In this method
the time-dependent densities of a TDHF collision slightly above the barrier were used
to minimize the energy, thus providing an impression of the underlying potential barrier between the nuclei.
The advantage of this method to the methods employing frozen nuclear densities, including,
the frozen Hartree-Fock (FHF)~\cite{denisov2002,washiyama2008,simenel2008,simenel2012}
and the density constrained FHF (DCFHF)~\cite{simenel2017} methods is that,
due to the dynamically changing density, it can account for dynamical rearrangements,
collective excitations at the mean-field level and in particular
couplings to vibrational~\cite{flocard1981,simenel2013a,simenel2013b} and 
rotational modes~\cite{simenel2004,umar2006d}, and the formation of a neck and
multi-nucleon transfer through the neck~\cite{simenel2008,umar2008a,simenel2010,sekizawa2013,vophuoc2016,godbey2017,jiang2020}.
In addition, the TDHF evolution and
the density-constrained minimization fully account for the Pauli principle~\cite{simenel2017a,umar2021}.
Microscopically, the effects of the Pauli principle on the nucleus-nucleus potential can also
be observed at the single-particle level, such as the splitting of orbitals with some
states contributing attractively (bounding) and some repulsively (antibounding) to the potential~\cite{umar2012b}.

Computationally, the implementation of the DC-TDHF method is similar to the
procedure used in the DCFHF approach except the constrained densities are
taken from the TDHF time-evolution at various times corresponding to the 
separation between the nuclear centers, R(t),
\begin{eqnarray*}
E_{\mathrm{DC-TDHF}}(t)&=&\underset{\rho}{\min}\left\{E[\rho_n,\rho_p]+\right.\nonumber\\
&&\sum_{q=p,n}\int\, d\mathbf{r} \ \lambda_q(\mathbf{r})  \left.\left[\rho_{q}(\mathbf{r})-\rho_{q}^{\mathrm{TDHF}}(\mathbf{r},t)\right]\right\}\,.
\end{eqnarray*}
The potential is then obtained by removing the binding energy of the HF ground-states,
\begin{equation}
V_{\mathrm{DC-TDHF}}(R)=E_{\mathrm{DC-TDHF}}(R)-E[\rho_1]-E[\rho_2]\,.
\label{eq:vrdctdhf}
\end{equation}
This barrier is then used to calculate fusion cross-sections via the method of
incoming wave boundary conditions, also used in the traditional coupled-channel calculations~\cite{hagino1999}.

Since the method only depends on the chosen EDF and no other parameters, the uncertainties
in the EDF parameters directly manifest themselves in the fusion cross-sections.
It should be noted also that, due to their dynamical nature, density rearrangements naturally depend on the 
energy of the collision, inducing an energy dependence to the nucleus-nucleus potential~\cite{washiyama2008,umar2009a,keser2012,umar2014a}.
Nevertheless, near and sub-barrier fusion cross-sections only require, within the DC-TDHF approach, one TDHF evolution at near-barrier central collision, showing overall good agreement with experiment~\cite{umar2009b,oberacker2013,simenel2013a,jiang2014,umar2014a,guo2018b,scamps2019b,godbey2019b}.
In this work, we then focus on sub and near-barrier fusion cross-sections and do not account for possible energy dependence of the potential which could affect above barrier cross-sections. 

\paragraph{Computational details.} Simulations were performed on a three dimensional Cartesian grid with no symmetry assumptions using
the program detailed in Ref.~\cite{umar2006c}.
The three-dimensional Poisson equation for the Coulomb potential
is solved by using Fast-Fourier Transform techniques
and the Slater approximation is used for the Coulomb exchange term.
The static solutions for the initial conditions are obtained through the damped gradient
iteration method~\cite{bottcher1989}. The box size used for all the calcium systems
was $44\times 28\times 28$~fm$^3$, whereas the oxygen and lead simulations was performed in a $42\times 24\times 24$~fm$^3$, with a mesh spacing of
$1.0$~fm used for all calculations. 
These values provide very accurate results due to the employment 
of sophisticated discretization techniques~\cite{umar1991a,umar1991b}.
The calculations have been repeated for a set of 
100 EDFs that have been sampled from the Bayesian posterior distributions determined in Ref.~\cite{mcdonnell2015} for the UNEDF1 parametrization~\cite{kortelainen2014}.

\begin{figure}[!htb]
	\includegraphics*[width=8cm]{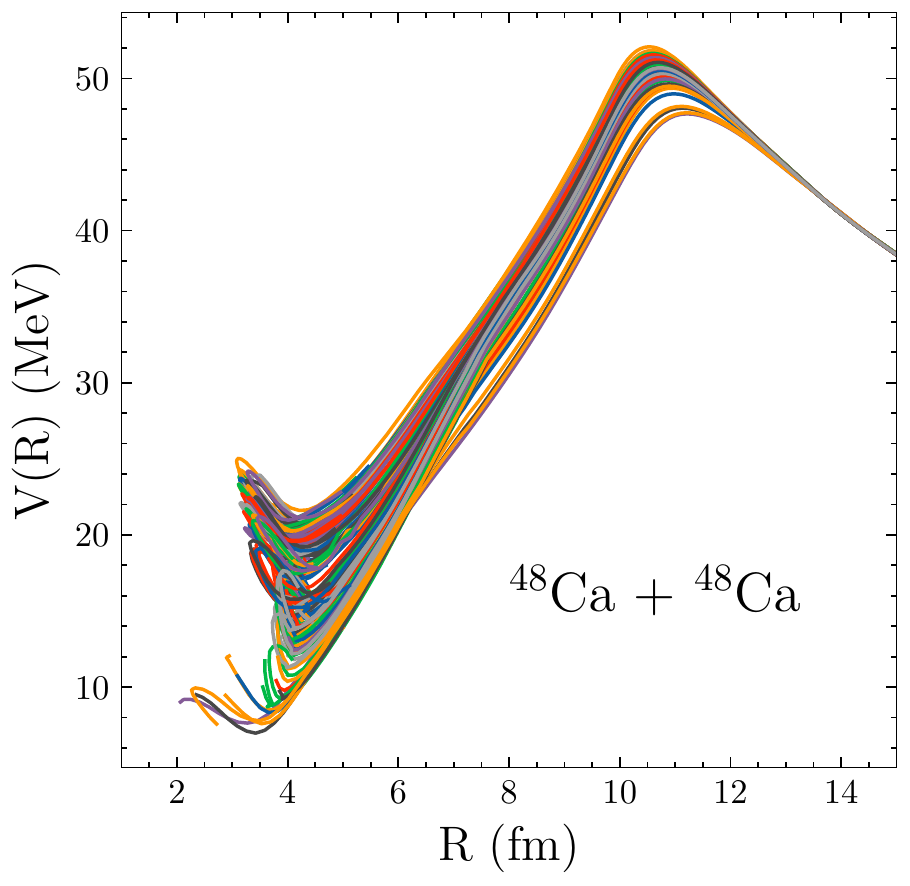}
	\caption{\protect Nucleus-nucleus interaction potentials for $^{48}\mathrm{Ca} + ^{48}\mathrm{Ca}$ at $\mathrm{E_{c.m.}}=56~\mathrm{MeV}$ for 100 EDFs sampled from the posterior distributions for UNEDF1.} 
	\label{fig:48Ca48CaPot}
\end{figure}

\paragraph{Fusion potentials.} An example of nucleus-nucleus potentials distribution is shown in Fig.~\ref{fig:48Ca48CaPot} for $^{48}$Ca$+^{48}$Ca. Each DC-TDHF potential is obtained from a TDHF calculation at $\mathrm{E_{c.m.}}=56$~MeV centre of mass energy, i.e., $\sim10\%$ above the barrier. 
Although all potentials exhibit similar behaviour, including inside the barrier, a striking observation is the resulting broad distribution of barrier heights from $V_B\simeq47.7$ to 52.1~MeV.
This $\sim8\%$ fluctuation in barrier height is clearly anti-correlated with the barrier radius that varies between $R_B\simeq10.5$ and $11.2$~fm. 

One can interpret these fluctuations in the barrier properties as an effect of the underlying variance in $^{48}$Ca neutron skin thickness that ranges from 0.13 to 0.22~fm in the static HF ground-state calculations with the same EDFs. 
Indeed, EDFs predicting the largest neutron thickness correspond to the largest values of $R_B$ and to the lowest values of $V_B$ (as the Coulomb repulsion decreases accordingly). 
The principal source of variance is in the relatively wide distribution of the slope of the nuclear matter symmetry energy, $L^{NM}_{sym}$, which directly effects the r.m.s. of the neutron distribution, and thus the neutron skin thickness of the nucleus.
Note that structure properties of excited states, such as low-lying quadrupole and octupole vibrations in $^{48}$Ca that could be affected by the EDF (see, e.g., Ref.~\cite{guo2018} for an investigation of the effect of tensor terms), could also impact the dynamical contribution (through coupling effects) to the nucleus-nucleus potential \cite{guo2018b,godbey2019c}. 

\begin{figure}[!htb]
	\includegraphics*[width=8cm]{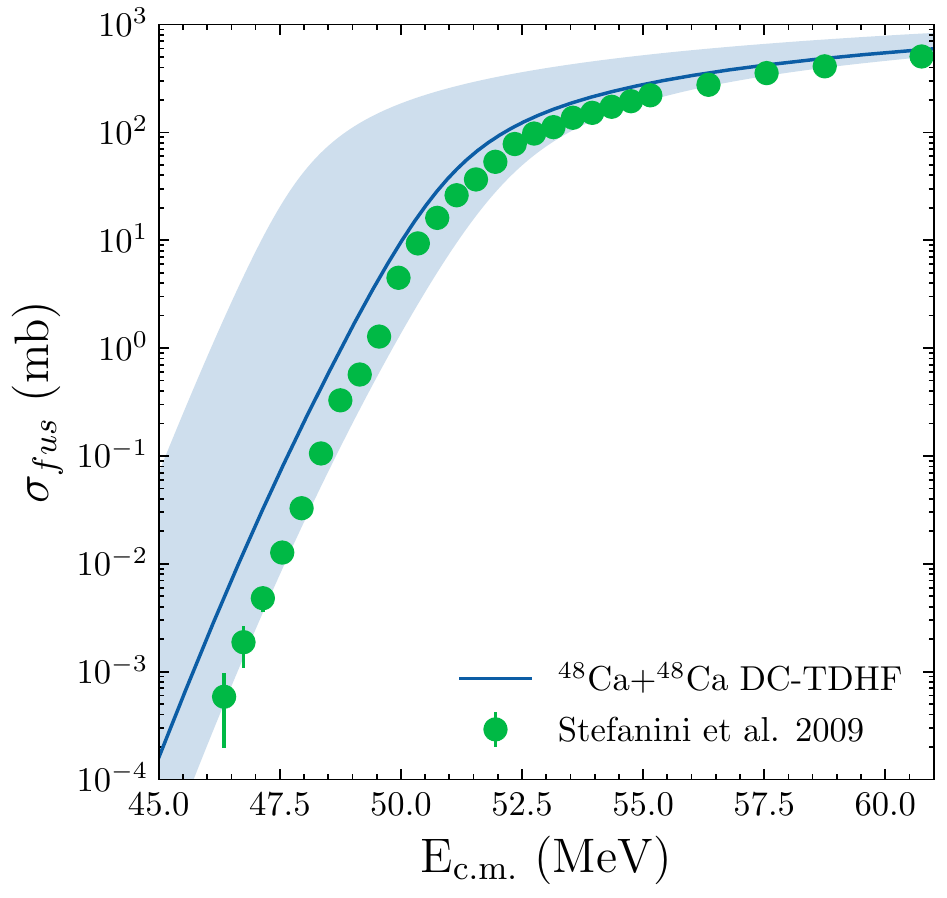}
	\caption{\protect Fusion cross-sections for $^{48}\mathrm{Ca} + ^{48}\mathrm{Ca}$. DC-TDHF median results are shown by the solid blue line with the light blue shaded area representing the uncertainty arising from the distribution of the EDF parameters. Experimental data is taken from~\cite{stefanini2009}.} 
	\label{fig:48Ca48CaCS}
\end{figure}

\paragraph{Fusion cross-sections.} Figure~\ref{fig:48Ca48CaCS} shows the predicted fusion cross-sections for $^{48}$Ca$+^{48}$Ca calculated with the potentials of Fig.~\ref{fig:48Ca48CaPot}. 
The associated uncertainty, shown by the light blue shaded area, arises from the underlying EDF parameters.
The solid blue line is the median of the predicted values and is not necessarily representative of the "optimal" results from the reported best-fit parameters of the EDF.
These cross-sections naturally exhibit large fluctuations at sub-barrier energies.
Indeed, the exponential dependence of  tunneling probability with  barrier height and width leads to variations of the predicted sub-barrier fusion cross-sections by up to $\sim3$ orders of magnitude.
This emphasizes the importance of constraining  quantities such as the symmetry energy in EDF determination.

\begin{figure}[!htb]
	\includegraphics*[width=8cm]{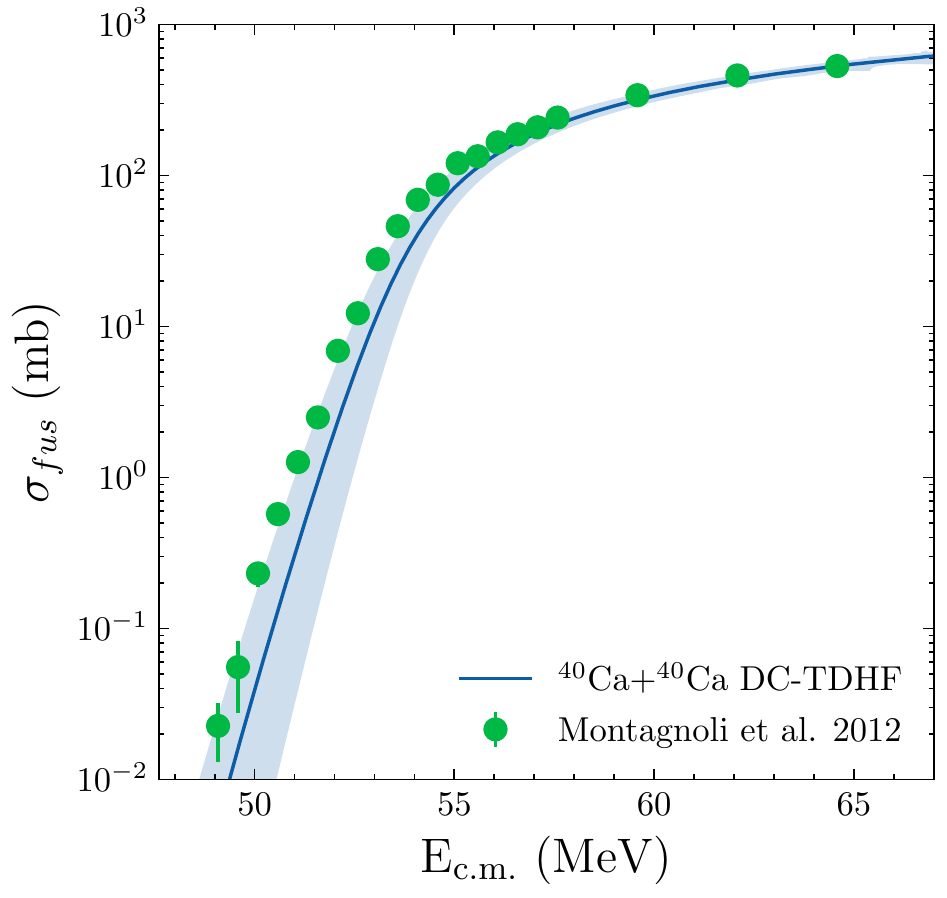}
	\caption{\protect Fusion cross-sections for $^{40}\mathrm{Ca}+^{40}\mathrm{Ca}$. DC-TDHF median results are shown by the solid blue line with the light blue shaded area representing the uncertainty arising from the distribution of the EDF parameters. Experimental data is taken from~\cite{montagnoli2012}.}
	\label{fig:40Ca40CaCS}
\end{figure}

The situation is much different in the $N=Z$ system $^{40}$Ca$+^{40}$Ca, as seen in Fig.~\ref{fig:40Ca40CaCS}.
Here, the uncertainties in fusion cross-sections are considerably reduced, though they still exceed an order of magnitude at lowest energy. 
This reflects smaller uncertainties in the nucleus-nucleus potential barrier properties, which in turn are associated with  small uncertainties in the density distribution of $^{40}$Ca ground-state.
One can note, however, that the sub-barrier slopes of the upper and lower confidence bands are different, leading to a increasing uncertainty as the energy goes down.
This effect is likely due to an impact of the varied EDFs on the nuclear shape dynamics inside the fusion potential, leading to an increase in uncertainty in the inner part of the potential.

\begin{figure}[!htb]
	\includegraphics*[width=8cm]{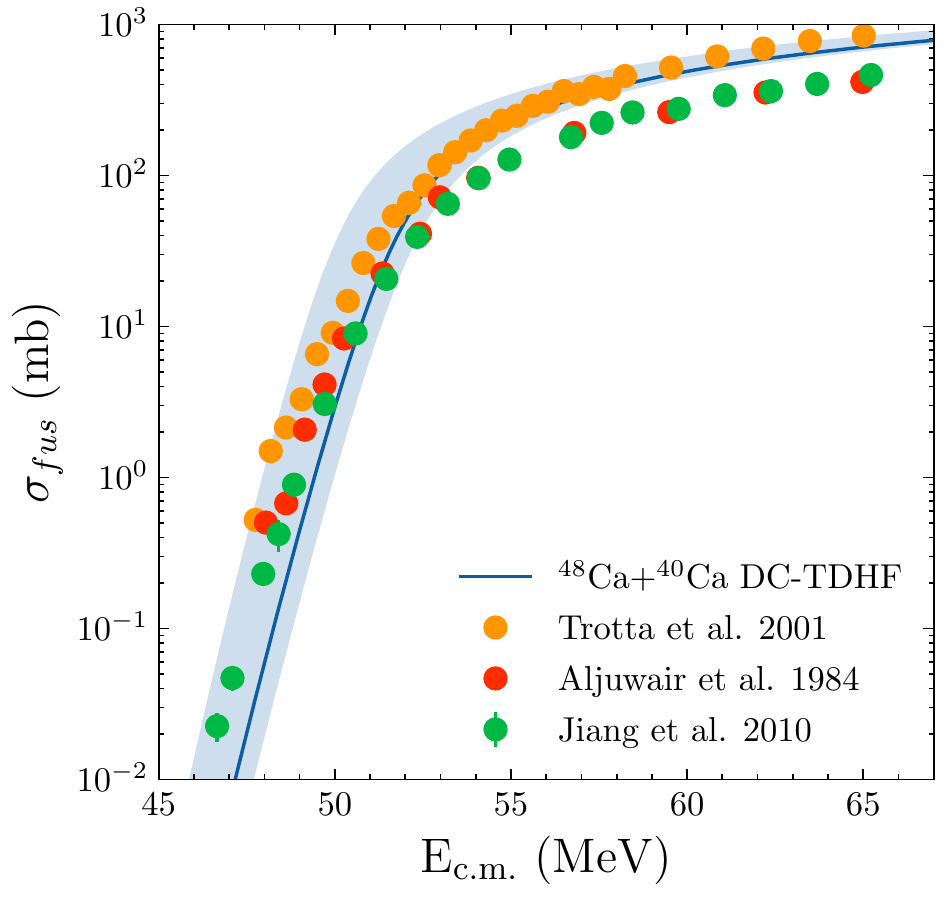}
	\caption{\protect Fusion cross-sections for $^{48}\mathrm{Ca} + ^{40}\mathrm{Ca}$. DC-TDHF median results are shown by the solid blue line with the light blue shaded area representing the uncertainty arising from the distribution of the EDF parameters. Experimental data is taken from~\cite{aljuwair1984,trotta2001,jiang2010}.} 
	\label{fig:48Ca40CaCS}
\end{figure}

\begin{figure}[!htb]
	\includegraphics*[width=8cm]{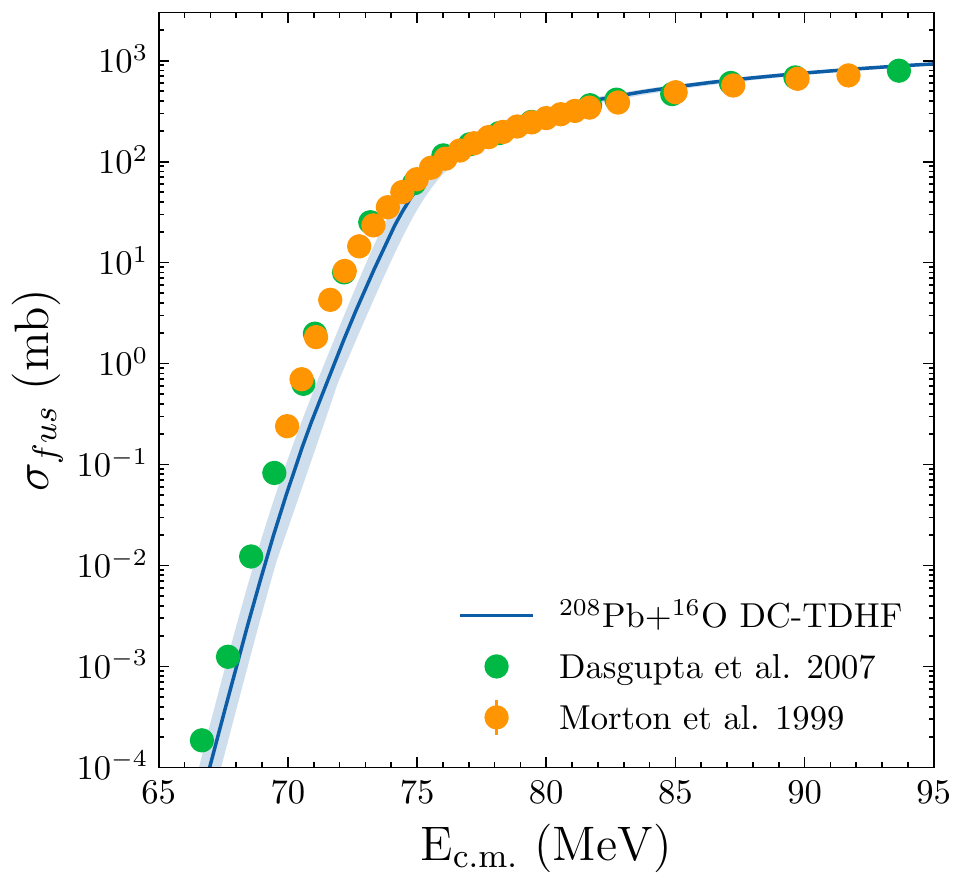}
	\caption{\protect Fusion cross-sections for $^{208}\mathrm{Pb} + ^{16}\mathrm{O}$. DC-TDHF median results are shown by the solid blue line with the light blue shaded area representing the uncertainty arising from the distribution of the EDF parameters. Experimental data is taken from~\cite{morton1999,dasgupta2007}.} 
	\label{fig:208Pb16OCS}
\end{figure}

We now consider reactions between asymmetric nuclei, namely $^{48}$Ca$+^{40}$Ca and $^{208}$Pb$+^{16}$O, which are both expected to undergo a net transfer of nucleons at contact due to a rapid charge equilibration process \cite{simenel2020}.
The resulting fusion cross-sections are plotted in Figs.~\ref{fig:48Ca40CaCS} and~\ref{fig:208Pb16OCS}.
Interestingly, no significant increase of uncertainties that could be attributed to transfer is observed. 
In fact, for $^{48}$Ca$+^{40}$Ca  a widening of the uncertainties in the sub-barrier fusion cross-sections similar to the $^{48}$Ca$+^{48}$Ca system is observed, though of smaller magnitude, which can be interpreted as a result of the uncertainty in $^{48}$Ca neutron skin thickness.

The observation of very small uncertainties  in $^{208}$Pb$+^{16}$O (see Fig.~\ref{fig:208Pb16OCS}) is somewhat surprising.
Indeed, the neutron skin thickness in $^{208}$Pb is found to fluctuate between $0.14$ and $0.19$~fm.
These fluctuations are of the same order as in $^{48}$Ca, and therefore similar uncertainties on fusion cross-sections would be expected. 
In fact, it has been shown that near-barrier fusion in $^{208}$Pb$+^{16}$O is triggered by a net exchange of $1-2$ protons from $^{16}$O to $^{208}$Pb \cite{simenel2008,simenel2010}.
A possible interpretation is that this dynamical effect does not fluctuate significantly with the EDF, and also washes out the static effects originating from the neutron skin. 
This interpretation is in agreement with earlier TDHF predictions that dynamical effects from charge equilibration have a stronger influence on  fusion barriers than static effects from neutron skins \cite{vophuoc2016}. 

Figures~\ref{fig:48Ca48CaCS}-\ref{fig:208Pb16OCS} also report experimental fusion cross-sections for comparison with the present calculations. 
Quantitative comparisons should be made with care, in particular near the barrier as the coupled channel effects are only treated at the mean-field level. 
Experimental results may also exhibit large fluctuations between data sets, as in $^{48}$Ca$+^{40}$Ca.
Nevertheless, one notes that, in going from neutron deficient to neutron rich calcium systems, one goes from an underprediction to an overprediction of sub-barrier fusion cross-sections by most EDFs. 
It would be interesting to see if including this trend in EDF fitting protocols would improve their overall qualities.

\paragraph{Conclusions.} Despite many advances in the rigorous uncertainty quantification of microscopic mean-field models for static properties of nuclei, there remain a number of features that are not well constrained by current approaches.
Properties such as the neutron radius of neutron-rich nuclei are particularly ill-constrained and contribute greatly to the total uncertainty arising in studies of reactions.
While this leads to a substantial amount of uncertainty arising from the model parameter determination, it also highlights the importance of including reaction data in the construction and constraint of new theoretical models that treat dynamics and structure on the same footing.
Such an approach has proven challenging, however, as the increased computational cost of including real-time simulations in the determination of model parameters is significant and precludes such a study even with modern computational facilities.
One promising method to make this feasible is to employ model order reduction techniques, such as the reduced basis method recently applied to static DFT~\cite{bonilla2022}.
Regardless of the technical specifics behind how new EDFs will be built, it is clear that the sensitivity of fusion reactions in particular to certain aspects of nuclear structure position it well as a probe for exotic nuclei.

\begin{acknowledgments}
This work has been supported by the U.S. Department of Energy under award numbers DE-SC0013847 (Vanderbilt University) and DE-NA0004074 (NNSA, the Stewardship Science Academic Alliances program) and by the Australian Research Council Discovery Project (project number DP190100256) funding schemes.
Portions of this research were conducted with the advanced computing resources provided by Texas A\&M High Performance Research Computing and the Institute for Cyber-Enabled Research at Michigan State University.

\end{acknowledgments}

\bibliography{VU_bibtex_master.bib}


\end{document}